\documentclass[twocolumn,notitlepage,aps,prb,10pt,superscriptaddress,floatfix,letterpaper,reprint]{revtex4-2}	

\usepackage{hyperref}
\usepackage[usenames,dvipsnames]{color}
\usepackage{amsmath}
\usepackage[english]{babel}
\usepackage{amssymb}
\usepackage{graphicx}
\usepackage{epstopdf}
\usepackage[normalem]{ulem}
\usepackage{subfigure}
\usepackage{braket}
\usepackage{bbold}
\usepackage{placeins}
\usepackage[para]{threeparttable}
\usepackage{lipsum}
\usepackage{multirow}
\usepackage{bm}
\usepackage{siunitx}

\usepackage{todonotes}

\definecolor{linkcolor}{HTML}{399B03}
\definecolor{urlcolor}{HTML}{399B03}
\hypersetup{pdfstartview=FitH, linkcolor=linkcolor,urlcolor=urlcolor,colorlinks=true}
\hypersetup{
    unicode=false,          
    pdftoolbar=true,        
    pdfmenubar=true,        
    pdffitwindow=false,     
    pdfstartview={FitH},    
    pdfauthor={},         
    colorlinks=true,        
    linkcolor=blue,         
    citecolor=red,          
}

\begin{document}

\title{Material specific optimization of Gaussian basis sets against plane wave data}
\author{Yanbing Zhou}%
\affiliation{%
 Department of Chemistry, University of Michigan, Ann Arbor, Michigan 48109, USA
}%
\author{Emanuel Gull}%
\affiliation{%
 Department of Physics, University of Michigan, Ann Arbor, Michigan 48109, USA
}%
\author{Dominika Zgid}%
\affiliation{%
 Department of Chemistry, University of Michigan, Ann Arbor, Michigan 48109, USA
}%
\affiliation{%
 Department of Physics, University of Michigan, Ann Arbor, Michigan 48109, USA
}%


\begin{abstract}

Since in periodic systems, a given element may be present in different spatial arrangements displaying vastly different physical and chemical properties, an elemental basis set that is independent of physical properties of materials may lead to significant simulation inaccuracies. To avoid such a lack of material specificity within a given basis set, we present a material-specific Gaussian basis optimization scheme for solids, which simultaneously minimizes the total energy of the system and optimizes the band energies when compared to the reference plane wave calculation while taking care of the overlap matrix condition number. To assess this basis set optimization scheme, we compare the quality of the Gaussian basis sets generated for diamond, graphite, and silicon via our method against the existing basis sets. The optimization scheme of this work has also been tested on the existing Gaussian basis sets for periodic systems such as MoS$_2$ and NiO yielding improved results. 

\end{abstract}

\maketitle

\section{Introduction}

In simulations of molecular systems, choosing a basis set for representing the electronic wave function is a well established procedure~\cite{Jensen_basis_set_revoew_2017,helgaker_bible,Gaussian-2,Gaussian-3,Gaussian-4}.
Gaussian basis sets that express a single orbital as a linear combination of primitive Gaussian orbitals are the most common molecular choice and multiple basis sets are available for most elements ~\cite{BSE_Schuchardt2007,BSE_Pritchard2019,Jensen_basis_set_revoew_2017, Gaussian-1}.

In simulations of periodic solids, many more choices of computational bases are commonly employed, including plane waves \cite{planeWave-1,planeWave-2} and linear augmented plane waves (LAPW) \cite{Wien2k_2020}. 
Over the years, density functional theory (DFT)\cite{DFT1,DFT2}  calculations of solids in the plane wave basis \cite{planeWave-1,planeWave-2} were far more commonly used than Gaussian \cite{DFTbasis,crystal} or any other basis set.
This is due to the sparsity of the 2-body integrals in the plane wave representation, as well as the systematic convergence of the basis with respect to the number of plane waves \cite{planeWave-3}.

However, plane wave calculations are only affordable due to the low computational scaling of DFT. Thousands of plane waves are necessary to reach an accuracy comparable to that of a Gaussian basis set \cite{planeWave-3,Sauer_pw_g_comp_2007} with many fewer basis functions. 
Consequently, when considering post-DFT, ab-initio calculations with a higher scaling than the $\mathcal{O}(n^3)$ scaling of DFT in the number $n$ of basis functions, Gaussian basis sets with fewer basis functions are an appealing choice. For instance, even simple post-DFT methods such  as the second order M\o ller-Plesset perturbation theory (MP2)~\cite{MP2-1,MP2-2,MP2-3,MP2-4,MP2-5,MP2-6,linearDependency-5} scale at least as $\mathcal{O}(n^5)$.

Moreover, when a physical or chemical interpretation of results is desired, Gaussian basis functions provide an invaluable tool to gain direct insight into the behavior of electrons in chemically relevant atomic orbitals. While similar insight can be obtained in a plane wave or LAPW calculation, it requires an additional projection procedure, for example to Wannier orbitals~\cite{Marzari12,LAPW}.
Finally, any calculation that requires the inclusion of core orbitals, such as the study of core ionization processes, requires an explicit representation of those orbitals as provided by Gaussian basis sets, rather than an effective treatment via pseudopotentials. 

While there are many advantages of employing a Gaussian basis for the description of solids, there are also clear drawbacks.
First, while there is a systematic way of improving a plane wave basis by adding additional orthogonal high-energy functions, this is not easily possible in a Gaussian basis, since the basis functions do not form an orthonormal set.
It is frequently observed that it is impractical or impossible to reach convergence with respect to the number of basis functions by simply adding additional Gaussian orbitals. This is due to linear dependencies that appear once a large number of Gaussians with small exponents (describing diffuse orbitals) are present~\cite{linearDependency-1,linearDependency-2,linearDependency-3,linearDependency-4,heyd05,linearDependency-5}.
Second, unlike in the case of plane waves, the exponents and coefficients of Gaussians need to be optimized for a given reference atomic problem. A minimization of the electronic energy with respect to a DFT calculation may optimize the basis function spanning the space of occupied orbitals (which determine the ground state energy) well, while leaving the ones necessary to express unoccupied orbitals relatively unoptimized.

Direct comparisons of results obtained with plane waves to those obtained with Gaussians are difficult, 
since plane wave bases rely on pseudopotentials, while Gaussian basis either explicitly treat core orbitals or employ different types of pseudopotentials. The differing description of core orbitals results in overall energy shifts of the remaining orbitals, making it somewhat cumbersome to compare the band structures between plane wave and Gaussian codes. Whether a differing band structure is a result of an insufficient Gaussian basis or of a different description of the core is often difficult to determine.

Despite of these disadvantages, the compactness of the basis and the ease of chemical/physical interpretation in the language of atomic orbitals are powerful advantages. However, the overall availability of Gaussian basis sets for solids is very limited, and the number of published basis sets for solids is a small fraction of the basis sets available for molecular systems \cite{pob12,pob19,pob18,lorenzo,morales}.
This is especially true for the 5th row transition metals such as  lanthanides, and actinides that are common in many newly synthesized  interesting compounds. It motivates our work on revisiting the  basis set optimization for solids. 

One of the difficulties in choosing a good Gaussian basis for the description of solids is that the same element may  display vastly different physical and chemical properties depending on its surroundings. For instance, different arrangements of carbon atoms create graphite, diamond, and graphene crystal lattices with both insulating and metallic character, respectively. Capturing these different behaviors with a single generic basis set, as is done in molecular systems, is difficult. We therefore explore a different paradigm in this work: rather than generating generic basis sets for solids, we optimize our basis sets in a material specific way (see also \cite{lorenzo} for related ideas).

The ground state energy remains an important target for the optimization. In addition, we minimize deviations from the orbital eigenvalues of a reference plane-wave calculation, while keeping basis functions as linearly independent as possible.
Using the orbital eigenvalues of a reference calculation as an optimization target has two effects. First, it avoids getting trapped in local energy minima which occur in the optimization of deep-lying orbitals. Second, it allows to also optimize the unoccupied orbitals, which do not directly enter the expressions for the ground state energy.

The material specific optimization approach retains the obvious advantage of Gaussian orbitals in that it produces compact bases with straightforward physical/chemical interpretation that can then be used in beyond-DFT calculations with higher-scaling methods. At the same time, by estimating the discrepancies between a Gaussian basis set and a non-Gaussian reference calculation, it eliminates the main disadvantage of working in a potentially incomplete basis set, and allows to quantify basis set errors.
We stress that this procedure is general and other reference calculations such as numerical bases or LAPWs may be used instead of plane waves.

In the reminder of this paper, we describe the optimization algorithm (Sec.~\ref{sec:method}), discuss results for commonly studied solids (Sec.~\ref{sec:results}), and present conclusions in Sec.~\ref{sec:conclusions}.

\begin{figure}[t!]
  \includegraphics[width=\columnwidth]{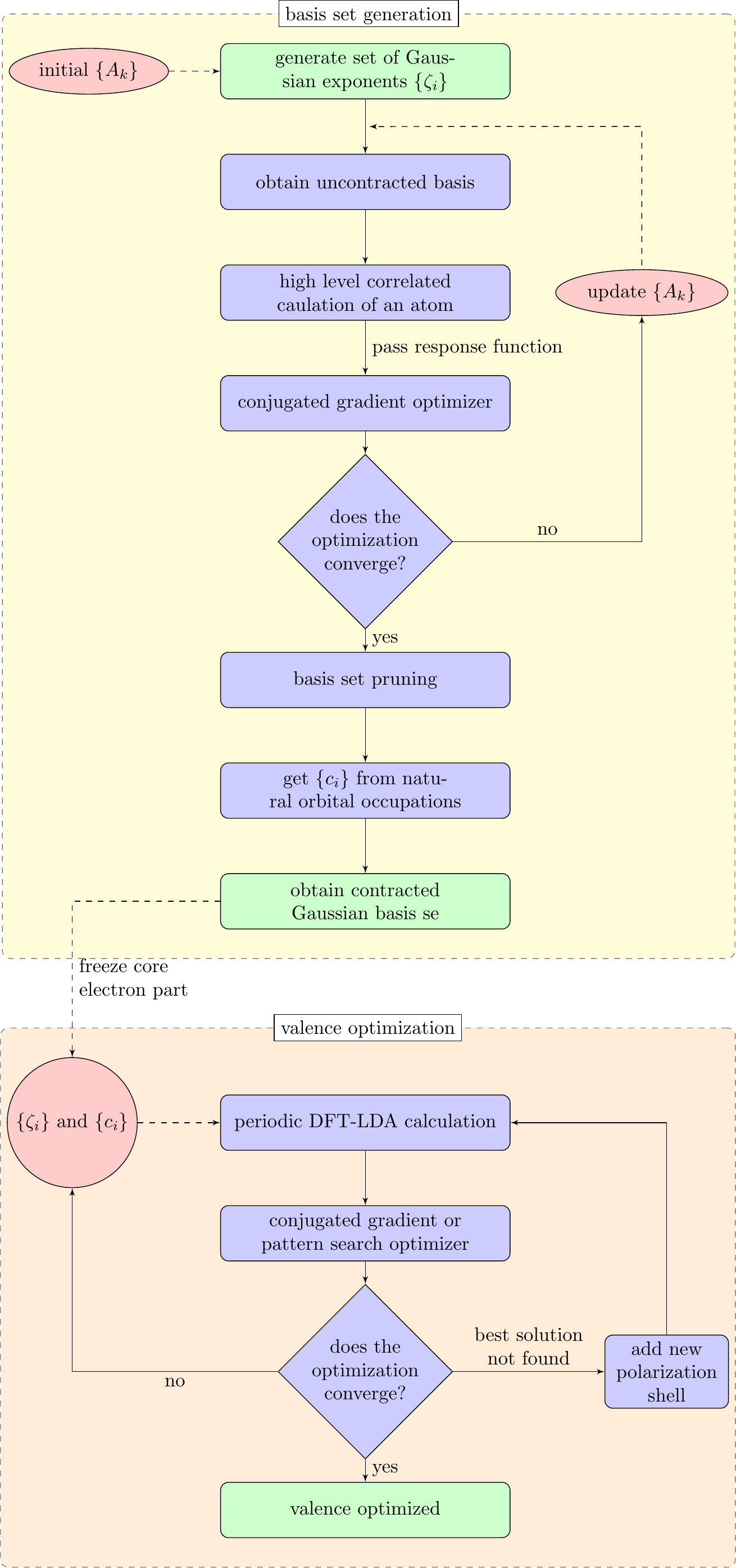}
  \caption{The basis set optimization scheme. Upper block: algorithmic steps for the generation of the initial exponents and contraction coefficients of the basis set. Lower block: steps for material-specific optimization of the valence orbitals.}
  \label{fig:flowchart}
\end{figure}

\section{Method}\label{sec:method}
Our basis set construction consists of two parts: {\bf (I)} {\em an atom-specific but materials-independent} generation of exponents and coefficients for an initial Gaussian basis set, followed by {\bf (II)} the optimization of the initial basis set in {\em a material specific} way for a given compound. In this section, we introduce Gaussian basis functions in ~\ref{ssec:Gaussian}, describe the initial optimization of the atomic problem in ~\ref{ssec:methodA}, and describe the subsequent material-specific optimization of the valence states in ~\ref{ssec:methodB}. The procedure is illustrated in Fig.~\ref{fig:flowchart}.

\subsection{Gaussian Orbitals} \label{ssec:Gaussian}
A basis set is defined as a set of single-particle functions (usually non-orthogonal atomic orbitals) that are then employed to build molecular orbitals ~\cite{Slater}.  In Gaussian basis sets, contracted Gaussian-type orbitals (cGTOs) are used to approximate atomic orbitals. The expansion of a single cGTO employs a linear combination of primitive Gaussian-type orbitals (GTOs)~\cite{IntroToChem,Helgaker}
\begin{equation}\label{eq:CGTO}
    \Phi^{cGTO}(x,y,z) = N\sum^{N_\text{prim}}_{i=1}{c_i \Phi_i^{GTO}(x,y,z)},
\end{equation}
where $c_i$ is the contraction coefficient, $N_\text{prim}$ describes the number of primitive GTOs, and $N$ is a normalization constant. A primitive GTO, $\Phi^{GTO}(x,y,z)$ is expressed in terms of spatial coordinates $x,y,$ and $z$ and integer exponents $a, b,$ and $c$ as 
\begin{equation}\label{eq:GTO}
    \Phi^{GTO}(x,y,z) = \tilde{N} x^a y^b z^c e^{-\zeta r^2},
\end{equation}
where $a+b+c$ controls the angular momentum quantum number, $\zeta$ is the exponent of the GTO controlling the width of the orbital, and $\tilde{N}$ is a normalization constant. Large $\zeta$ results in a 
tight function while small $\zeta$  gives a diffuse function.

\subsection{Generation of Initial Basis Set}\label{ssec:methodA}
The first step of our procedure (top panel of Fig.~\ref{fig:flowchart}) generates exponents and contraction coefficients for each of the atoms. This procedure is not material specific and  it is primarily used to optimize the cGTOs for the atomic core.

While the basis set generation procedure detailed above is not new, we describe it here for the sake of completeness. For many elements, it may be possible to start from existing atomic basis sets. However, for ``more exotic'' elements such as rare earth metals a generation of the atomic basis may be either necessary or desired.

\subsubsection{Generation of starting exponents}\label{sec:basis_generation}
To find an initial set of exponents $\zeta_i$ for the primitive Gaussians $\Phi^{GTO}_i$ that will later be assembled into cGTOs, we follow the procedure described in Ref.~\cite{Legendre}. There, the natural logarithm of each of the exponents $\zeta_i$, $i=1,\dots,N_{\rm prim}$ is expanded into an orthonormal Legendre polynomial expansion 
\begin{equation}\label{eq:Legendre}
    \ln{\zeta_i} = \sum^{k_{\rm max}}_{k=0} A_k P_k(\frac{2i-2}{N_{\rm prim} -1} -1),
\end{equation}
in terms of coefficients $A_k$ which are common for different $\zeta_i$ exponents. $N_{\rm prim}$ is the number of Gaussian primitives in a given contracted orbital, $P_k$ is the Legendre polynomial of order $k$. The initial set of parameters $A_k$ are  obtained according to the procedure described in Ref.~\cite{Legendre}. In our work, we set $k_{\rm max} = N_\text{prim} - 1$  or $k_{\rm max} = 6$, whichever is greater. 
The expansion into a Legendre basis eliminates the problem of linear dependencies that occurs when initial values of the exponents are optimized independently.

\subsubsection{Initial optimization of exponents and coefficients } \label{sssec:coeff}
With the initial set of $\{\zeta_i\}$ obtained, a first uncontracted atomic basis set is generated with a total number of $k_{max}+1$ variational parameters $A_k$. 
This uncontracted basis is employed as a starting guess for the energy minimization in an atomic Hartee-Fock (HF) calculation.
We use  a conjugate gradient minimization~\cite{Hestenes52,dakota} scheme to find the variational parameters $\{A_k\}$ that minimize the electronic energy of the atom. Subsequently, the atom parameters are re-optimized in the configuration interaction singles and doubles (CISD) method.

Once the optimization of the $\zeta_i$ converges, contraction coefficients $c_i$ for cGTOs can be found. They are generated as described in Ref.~\cite{contraction} by obtaining the expansion coefficients of the natural orbitals obtained by CISD  in the basis of primitive GTOs. 
After this step, the number of cGTOs is then equal to the number of uncontracted functions.

\subsubsection{Basis set pruning}\label{sssec:methodA3}
Based on the natural orbital occupations of the CISD solution, many cGTOs typically have occupancies close to zero.  We eliminate these cGTOs from the calculation by choosing to retain only the ones with an occupation threshold greater than 0.001.

In addition, diffuse Gaussian primitives with exponents smaller than 0.1 are eliminated, in analogy to Refs.~\cite{dovesi90,pob12,pob19}. This approximation was found to remove numerical instabilities in the calculation of solids.
The resulting atomic orbitals can then be separated into core and valence orbitals.

\subsection{Valence Optimization}\label{ssec:methodB}
The set of exponents and contraction coefficients obtained from Sec.~\ref{sssec:coeff} and pruned in Sec.~\ref{sssec:methodA3} form the starting point of the valence optimization. 
Unlike the calculations in Sec.~\ref{sssec:coeff} and Sec.~\ref{sssec:methodA3}, this procedure is material specific and adjusts the valence orbitals to the chemical environment present  in a given solid, while the core configuration is assumed to be material-independent and remains frozen.

A basis optimization where only the total energy $E_{tot}$ is minimized has the tendency to optimizes only the lowest eigenvalues well. 
However, the optimization of higher occupied orbitals may then become more difficult, since they contribute only little to the lowering of the overall DFT total electronic energy which is partially proportional to the sum of occupied orbital eigenvalues. As a consequence, the optimization may get stuck in a local minimum. 

To remedy this issue, we choose an additional optimization criterion based on the difference between the molecular orbital (MO) energy at each k-point between the Gaussian calculation and a reference calculation. In our case, the reference is derived from a plane-wave calculation of the same system, taking advantage of the fact that the quality of the plane-wave basis set can be systematically controlled by a single parameter, the cut-off energy ($e_{cut}$), and can be converged with respect to the basis set size.
Note, however that different pseudopotential present in the Gaussian and/or the plane wave calculation may lead to an overall shift of all energy bands. This shift is captured by an overall orbital-independent but potentially k-dependent energy shift parameter $\Lambda(k)$. The procedure of finding $\Lambda(k)$ is explained in Sec.~\ref{sssection:optimizaEigenV}.

The minimization of the deviation from a reference band structure is characterized by ${\Vert G(\{\zeta_i, c_i\}) - P(e_{cut}) - \Lambda(k) \Vert}_{F}$, where $G(\{\zeta_i, c_i\})$ are MO eigenvalues of the Gaussian basis set calculation at each k-point, $P(e_{cut})$ are the corresponding plane-wave eigenvalues, and $\Lambda(k)$ is a k-dependent shift value. $||\cdot||_F$ denotes the Frobenius norm. 

A third optimization criterion is given by the condition numbers of the overlap matrices at each k-point, which are minimized to obtain as little linear dependence in the resulting basis as possible. In a nearly linearly dependent Gaussian basis, the condition number can be very large or even infinite. Acceptable conditions numbers usually are smaller than $10^{5}$. Consequently, denoting the vector of condition numbers of the overlap matrices  as $\vec{\kappa}$, the minimization of the term $\gamma_{2} \Vert \vec{\kappa}(\{\zeta_i, c_i\}\Vert_{2})$, with $\Vert \cdot \Vert_2$ denoting the Euclidian norm, provides a penalty for linearly dependent solutions. Constraints on the condition number of the overlap matrix are frequently used in this context, see {\it e.g.} Refs.~\onlinecite{ovlpConstrain-1, ovlpConstrain-2, ovlpConstrain-3, lorenzo}.

These three conditions lead to a minimization problem of the valence orbitals for a set of exponents of the Gaussian primitives and corresponding contraction coefficients $\{ \zeta_i, c_i\}$ that minimize the functional
\begin{align}\label{eq:optfn}
&\Omega(\{\zeta_i, c_i\}) = E_{tot}(\{\zeta_i, c_i\})+ \nonumber \\
&+ \gamma_{1}{\Vert G(\{\zeta_i, c_i\}) - P(e_{cut}) - \Lambda(k) \Vert}_{F} \\
&+ \gamma_{2} {\Vert \vec{\kappa}(\{\zeta_i, c_i\})\Vert}_{2},\nonumber 
\end{align}
where $\gamma_{1}$ and $\gamma_{2}$ are weight factors that control the relative contributions of each of the three terms. In this work, $\gamma_{1}$  was set between $100$ and $10^{3}$ and $\gamma_{2}$ between $10^{-5}$ and $1$. 

In practice, we minimize the functional $\Omega(\{\zeta_i, c_i\})$ with a combination of a conjugate gradient optimizer and a non-gradient based pattern search~\cite{pattern_search} optimization which we use to overcome local minima. 

The bottom half of Fig.~\ref{fig:flowchart} illustrates the workflow for optimizing the basis functions for the valence orbitals. After convergence of the valence optimization, the core orbitals can be selectively re-optimized. 

Note that the valence optimization method described here can also be applied to established basis sets, adapting a general-purpose basis set to a specific material context.

\subsubsection{The value of \texorpdfstring{$\Lambda$}{Lambda} in the eigenvalue optimization}\label{sssection:optimizaEigenV}

A systematic shift $\Lambda(k)$ is frequently present between the plane wave and Gaussian basis set codes. This quantity frequently makes the comparisons of band diagrams between the plane wave codes and Gaussian orbital codes inconvenient. 
While in general the value of $\Lambda$ is k-dependent, in practice since the core orbitals are well localized they form flat and approximately k-independent energy bands. This is why in this work, we assume that $\Lambda$ is k-independent.

The vale of $\Lambda(k)$ can be evaluated by noticing several facts. First, we assume that the pseudopotential in the converged plane wave calculations was created to account for the low lying occupied orbitals and it recovers the energies of low lying occupied orbitals very accurately. \cite{pseudo1,pseudo2}
Then an all-electron calculation or a calculation with an accurate pseudopotential in an infinite/optimal Gaussian basis set can yield eigenvalues that are different only by a constant factor that is result of a freedom to add a constant shift in the pseudopotentials employed during the plane wave calculation.
Consequently, for every k-point, in an optimized bases the value of the shift $\Lambda$  can be found using the least square fit between eigenvalues $p_j$ in the plane wave basis and $g_j$ Gaussian basis $\min \sum^{N}_{j}  {(p_j - g_j - \Lambda)}^2$. We can find the shift $\Lambda$ by finding the minimum of the above equation as a function of $\Lambda$, 
$\frac{\partial}{\partial \Lambda} (\sum^{N}_{j}  {(p_j - g_j - \Lambda)}^2) = 0$. Finally, the value of the $\Lambda$ shift is given by
\begin{equation}\label{eq:shift}
	\Lambda = \sum^{N}_{j} \frac{(p_j - g_j)}{2N}.
\end{equation}

The knowledge of the value of  $\Lambda$ allows us to use the constrain ${\Vert G(\{\zeta_i, c_i\}) - P(e_{cut}) - \Lambda(k) \Vert}_{F}$ helping the optimization of eigenvalues and to match approximately (within the desired criteria) the occupied eigenvalues. It also can be used to optimize and match a selected number of unoccupied eigenvalues. In this way, the quality of the Gaussian basis is rather systematically improved since we can choose and systematically improve how many unoccupied eigenvalues are being optimized. We can also observe which groups of AO orbitals have to be added to the basis to match increasingly higher lying virtual eigenvalues.
Moreover, finding the value of the shift described above, allows us to easily examine the quality of Gaussian basis sets prior to their usage by comparing the results to a converged plane wave calculation. Note also the even if a user does not desire to use such a specific constrain relying on matching the eigenvalues, it can be used to drive the optimization in the initial optimization steps and subsequently can be released allowing us to only insist on the optimization of the value of the total energy during the later steps.

\section{Results}\label{sec:results}
In this section, we assess the generation and optimization scheme described in the previous sections. Our discussion consists of two parts. In Sec.~\ref{sec:basis_gen_C_SI}, we showcase the accuracy of the bases generated for simple monoatomic solids such as diamond, graphite, and silicon. We then use the optimization scheme developed in Sec.~\ref{ssec:methodB} to optimize the existing basis sets pob-DZVP-rev2 and pob-TZVP-rev2 in a material specific way for the solids BN, MnO, MoS$_2$, NiO. We also to assess the quality of existing GTH basis sets. 

Unless otherwise mentioned, all the calculations presented here using Gaussian basis sets are performed with CRYSTAL17 \cite{crystal17} using DFT with the local density approximation (LDA). The reference calculations were performed in a plane wave basis in GPAW \cite{gpaw} with an energy cutoff of 2000 eV and a Monkhorst-Pack grid of $8 \times{8}\times{8}$ k-points.
We use the DAKOTA \cite{dakota} conmin-frcg DAKOTA coliny-pattern-search package as optimizers.
For all cases analyzed, the optimizer converges within 100 optimization steps. 

\subsection{Basis generation for diamond, graphite, and silicon}\label{sec:basis_gen_C_SI}
This basis generation scheme is intended to be material specific and a different basis set should be generated for every compound examined. While some of these bases for a given atom may be transferable between different materials, we generally believe that due to significantly different properties of solids containing the same atoms (eg. diamond, graphite, and graphene), a materials specific basis optimization is preferred in calculations of solids.

The initial atomic basis sets for C and Si are generated using the method described in Sec.~\ref{sec:basis_generation} and schematically presented in the yellow shaded block of Fig.~\ref{fig:flowchart}.
Note that the high level method used to perform correlated calculations in an uncontracted basis for C and Si atoms is configuration interaction singles and doubles (CISD) from Psi4 \cite{psi4}. This step allows us to find the initial set of exponents and contraction coefficients that are then pruned, taking advantage of natural orbitals. 
Subsequently, each shell present in the core part of the basis is first optimized individually using the functional from Eq.~\ref{eq:optfn}. The resulting core basis functions are then frozen and the valence basis functions are optimized by the conjugate gradient optimizer, with a pattern search optimizer as a further refinement tool. This optimization procedure can be done for the atom or in a material specific way, where the valence part of the basis is then reoptimized again for each of the compounds.

\begin{figure}[tbh]
\includegraphics[width=\columnwidth]{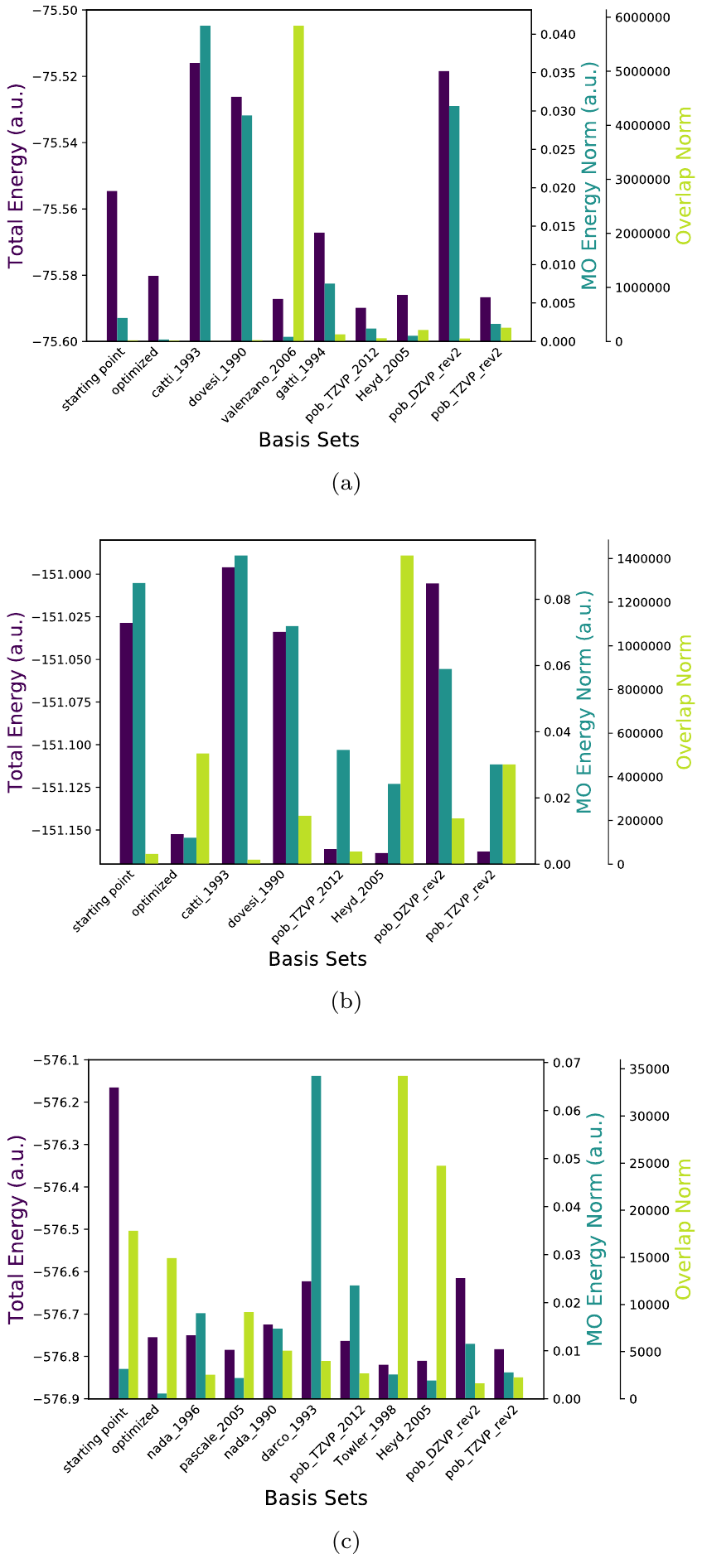}%
\caption{A comparison of generated bases with existing Gaussian bases for :(a) diamond;(b)  graphite;(c)  silicon.}%
\label{fig:basis_generation}%
\end{figure}

To examine the quality of the generated basis sets, we compare them to a list of existing basis sets for solid state calculations~\cite{cryst_basis_lib} of diamond, graphite, and silicon.
In Fig.~\ref{fig:basis_generation} we show the results of the comparison between the basis set obtained from the atomic level optimization, the material specific optimized basis set, and the existing basis sets listed in literature.

For diamond (see Fig.~\ref{fig:basis_generation} panel a)) the basis set generated has 4 $s$ shells, 3 $p$ shells, and 1 $d$ shell. The number of primitives in each shell is illustrated in Tab.~\ref{tab:shell_primitives_diamond}.

We observe that this initial atomic basis set, before the material specific optimization, has moderately high electronic energy comparable to older available basis sets such as the one optimized by Gatti~\cite{gatti94} in 1994. However, both its energy and its MO norm are much lower than pob-DZVP-rev2~\cite{pob19}. The optimized basis is comparable to pob-TZVP-rev2~\cite{pob19}  and the original pob-TZVP~\cite{pob12} basis as well as the basis developed by Heyd~\cite{heyd05}. While the electronic energy in these bases is slightly lower, the overlap norm and the MO eigenvalue norm are higher.

For graphite (see Fig.~\ref{fig:basis_generation} panel b)), the basis set generated has the same shell structure as the diamond basis, while the number of Gaussian primitives are slightly different, as shown in Tab.~\ref{tab:shell_primitives_graphite}. 
The optimized basis is again comparable in its quality to the pob-TZVP-rev2~\cite{pob19}. However, it has a much lower MO norm while the norm of the overlap remains comparable. 

For silicon (see Fig.~\ref{fig:basis_generation} panel c)), the basis set generated has 4 $s$ shells, 3 $p$ shells, and 1 $d$ shell. The number of primitives in each shells is shown in Tab.~\ref{tab:shell_primitives_silicon}.
The optimized basis yields comparable results to the pob-TZVP bases. However, the optimized basis has a lower MO norm while having a slightly higher overlap norm. 

In general, the total energies of our basis sets are comparable to the pob-TZVP basis sets, while the MO energy norms are much lower and the overlap condition numbers remain reasonable. This is partly due to the fact that we are optimizing multiple parameters at once, i.e., optimizing both the eigenvalue norm and the total energy together, instead of minimizing the energy alone. Note that the MO constraint causes the equal optimization of all eigenvalues, independent of their magnitude. 
This means that we can design a basis to optimize a  particular set of eigenvalues of interest, either in the occupied part, or in the unoccupied part. We can also choose to optimize a certain number of bands in a specific energy window. 

It is also worth mentioning that, although the three terms in $\Omega(\{\zeta_i, c_i\})$ from Eq.~\ref{eq:optfn} may appear independent to each other, mutual dependencies between them are observed.
As demonstrated in Fig.~\ref{fig:basis_generation}, the MO norm and the total energy value varying patterns are consistent with each other independent of a specific  basis set. When the MO norm is high the total energy associated with it also tends to be higher, and once the MO norm is minimized the resulting energy is low. This trend is expected and confirms that MO norm is a good indicator of the basis set quality and can be used as a driving factor during the basis set optimization. 
The overlap norm anti-correlates with the value of the total energy. When the overlap norm is very small, the total energy tends to increase. This is expected, as the optimization of any basis has to balance orthogonality of the basis functions with the overall the lowering of the total energy.

\begin{center}
\begin{table}
\begin{tabular}{ p{2cm}| p{2cm} p{2cm} p{2cm}}
\hline
\hline
\hfil Shell Index & \hfil $s$-shell & \hfil $p$-shell & \hfil $d$-shell\\
\hline
\hfil $1$ & \hfil 6 & \hfil   & \hfil   \\
\hfil $2$ & \hfil 4 & \hfil 4 & \hfil   \\
\hfil $3$ & \hfil 1 & \hfil 1 & \hfil 1 \\
\hfil $4$ & \hfil 1 & \hfil 1 & \hfil   \\ 
\hline
\hline
\end{tabular}
\caption{Number of Gaussian primitives in each shell in the optimized C-diamond basis set.} \label{tab:shell_primitives_diamond}
\end{table}
\end{center}

\begin{center}
\begin{table}
\begin{tabular}{ p{2cm}| p{2cm} p{2cm} p{2cm}}
\hline
\hline
\hfil Shell Index & \hfil $s$-shell & \hfil $p$-shell & \hfil $d$-shell\\
\hline
\hfil $1$ & \hfil 6 & \hfil   & \hfil   \\
\hfil $2$ & \hfil 4 & \hfil 3 & \hfil   \\
\hfil $3$ & \hfil 1 & \hfil 1 & \hfil 1 \\
\hfil $4$ & \hfil 1 & \hfil 1 & \hfil   \\ 
\hline
\hline
\end{tabular}
\caption{Number of Gaussian primitives in each shell in the optimized C-graphite basis set.} \label{tab:shell_primitives_graphite}
\end{table}
\end{center}

\begin{center}
\begin{table}
\begin{tabular}{ p{2cm}| p{2cm} p{2cm} p{2cm}}
\hline
\hline
\hfil Shell Index & \hfil $s$-shell & \hfil $p$-shell & \hfil $d$-shell\\
\hline
\hfil $1$ & \hfil 8 & \hfil   & \hfil   \\
\hfil $2$ & \hfil 8 & \hfil 4 & \hfil   \\
\hfil $3$ & \hfil 1 & \hfil 2 & \hfil 1 \\
\hfil $4$ & \hfil 1 & \hfil 1 & \hfil   \\ 
\hfil $5$ & \hfil 1 & \hfil 1 & \hfil   \\
\hfil $6$ & \hfil 1 & \hfil 1 & \hfil   \\
\hline
\hline
\end{tabular}
\caption{Number of Gaussian primitives in each shell in the optimized Si basis set.} \label{tab:shell_primitives_silicon}
\end{table}
\end{center}

\subsection{Material specific optimization of existing bases}\label{sec:basis_optimization_existing_bases}

\subsubsection{\texorpdfstring{MoS$_2$}{MoS2}}\label{section:MoS2}

\begin{figure*}%
\includegraphics[width=\textwidth]{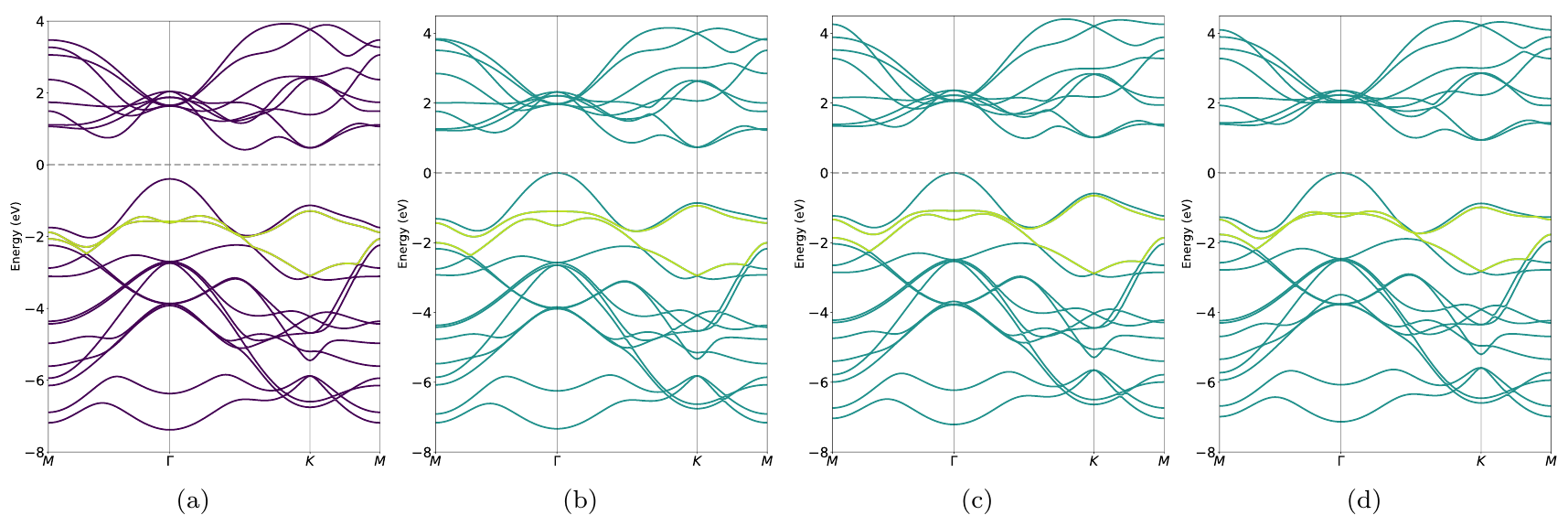}%
\caption{MoS$_2$ band structure plots:
(a) plane-wave basis calculation with cut off energy: 2000eV;
(b) Gaussian basis calculation with original pob-DZVP-rev2-basis;
(c) Gaussian basis calculation with original pob-TZVP-rev2-basis; and,
(d) Gaussian basis calculation with optimized pob-TZVP-rev2-basis.}%
\label{fig:MoS2}%
\end{figure*}

\begin{figure}
  \includegraphics[width=\columnwidth]{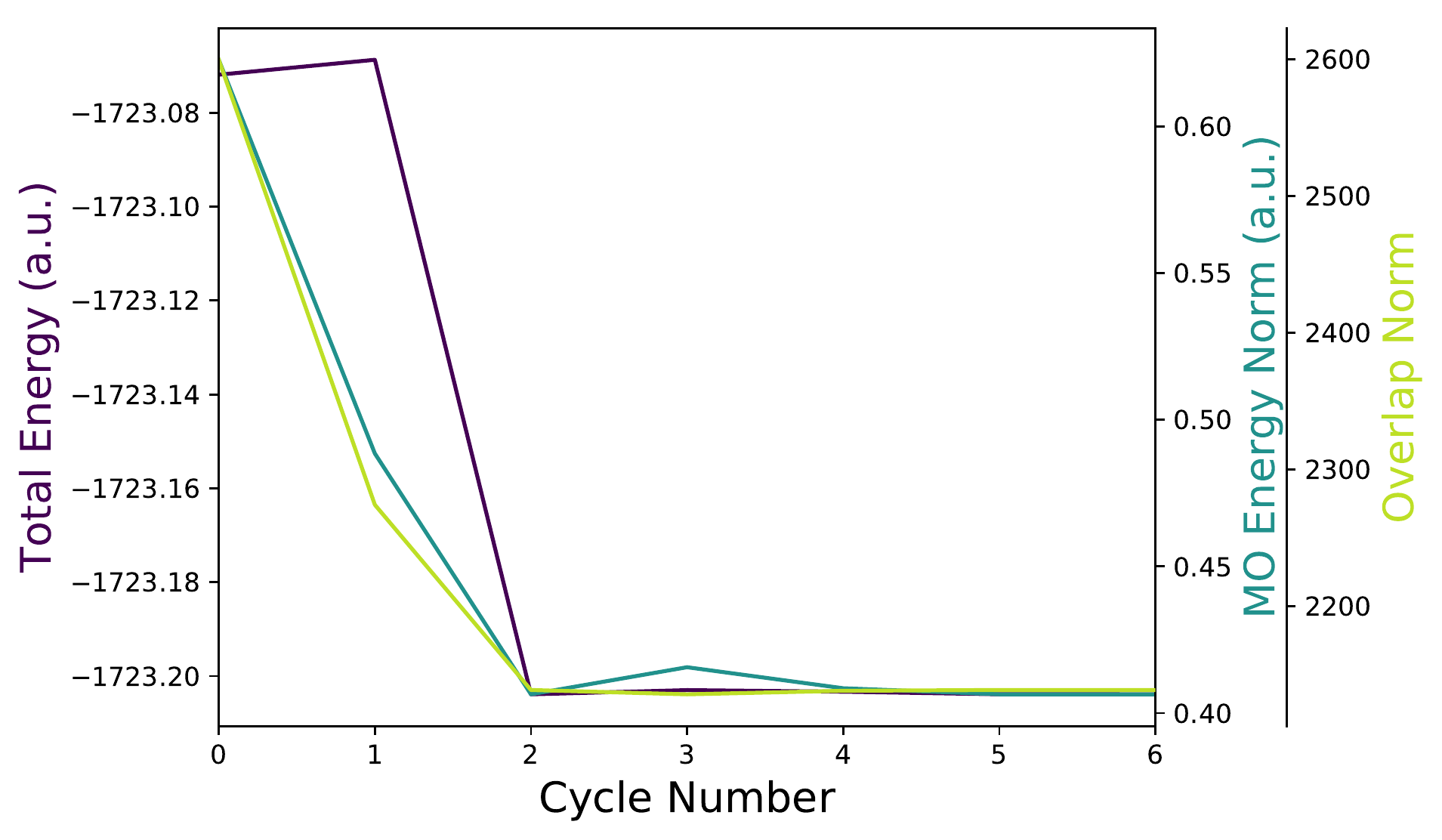}
  \caption{Values of the total energy (purple curve), MO energy norm (dark green curve), and overlap norm (light green curve) during the minimization of the functional $\Omega$ of Eq.~\ref{eq:optfn} for solid MoS$_2$. The minimization was performed using a conjugate gradient optimizer. For details see Sec.~\ref{section:MoS2}.}%
  \label{fig:cg}
\end{figure}

For MoS$_2$ (Pearson symbol: hP6, crystal system: hexagonal, space group number: 194) the calculations performed in the pob-DZVP-rev2 basis show that the high lying occupied band marked in light green and shown in Fig.~\ref{fig:MoS2} panel b) do not show the band crossings that are present in the reference plane waves calculations. The calculations conducted in the pob-TZVP-rev2 presented in Fig.~\ref{fig:MoS2} panel c) do not show a qualitative improvement when compared to pob-DZVP-rev2 results.
Optimization procedures following Sec.~\ref{sssec:methodA3} are performed with Mo's ECP, 4s, 4p and 4d shell frozen and S's 1s, 2s and 2p core frozen. 
We observe that with the existing number of shells in the basis set, the optimization of pob-DZVP-rev2 basis does not recover the band crossing present in the reference. 
For the pob-TZVP-rev2 basis, our optimization scheme helps to improve the predicted band structure, see Fig.~\ref{fig:MoS2} panel d), and is showing a consistently better band structure behavior when comparing to the reference result.
Fig.~\ref{fig:cg} shows the convergence of the material specific optimization process where we start with the pob-TZVP-rev2 basis. The values of the three individual terms of Eq.~\ref{eq:optfn} are displayed separately for each of the cycles. 

To showcase our results clearly, we list the three terms present in Eq.~\ref{eq:optfn} separately and examine the value of the total energy, the norm of the overlap matrix, and the difference in the MO eigenvalues for every k-point after accounting for the presence of the constant shift.

In the optimization process, we find that the Mo optimized pob-TZVP-rev2 is significantly different from its library version. The changes due to the optimization of the pob-TZVP-rev2 basis of S are less pronounced but also important.

\begin{figure*}%
\includegraphics[width=\textwidth]{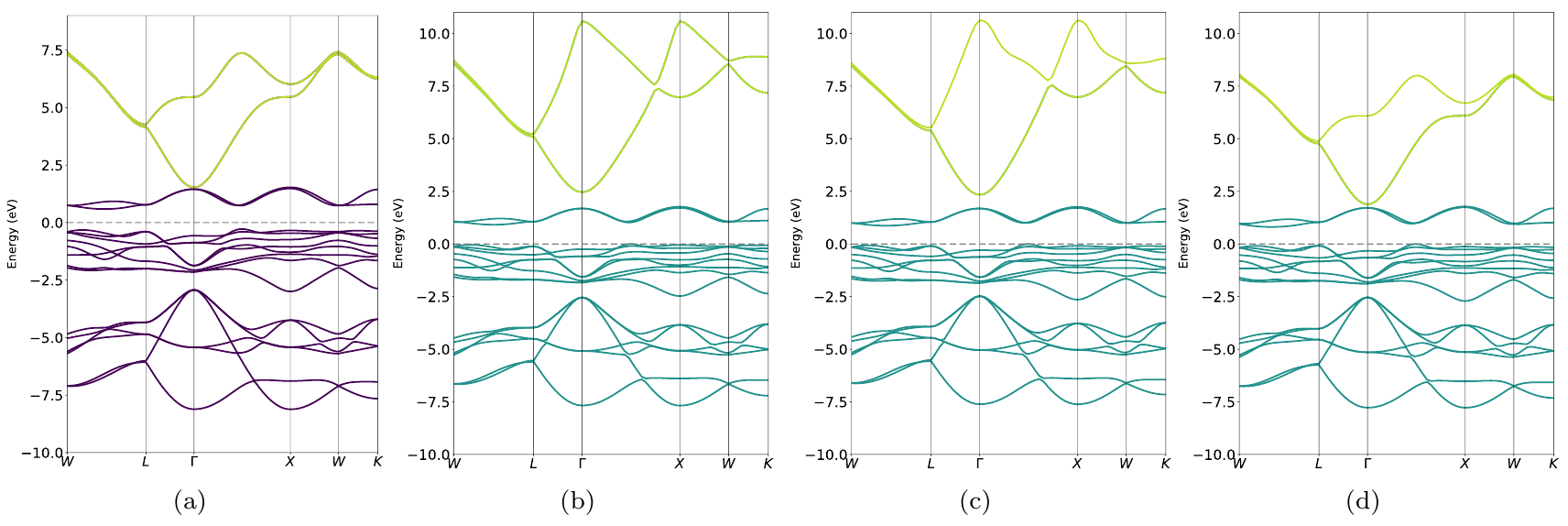}%
\caption{NiO band structure plots:
(a) plane-wave basis calculation with cut off energy: 2000eV;
(b) Gaussian basis calculation with original pob-DZVP-rev2-basis;
(c) Gaussian basis calculation with original pob-TZVP-rev2-basis; and,
(d) Gaussian basis calculation with pob-TZVP-rev2-basis and an additional d shell.}%
\label{fig:NiO}%
\end{figure*}

\subsubsection{NiO}

For NiO (Pearson symbol: cF8, crystal system: fcc rocksalt, space group number 225, AFM) the calculated band structure Fig.~\ref{fig:NiO} panel b) and Fig.~\ref{fig:NiO} panel c) using pob-DZVP-rev2 and pob-TZVP-rev2 basis,respectively are both qualitatively inconsistent with the reference Fig.~\ref{fig:NiO} panel a) for the unoccupied band marked in light green. Moreover, the  pob-TZVP-rev2 basis does not lead to any significant improvement over pob-DZVP-rev2. We attempted the optimization of both pob-DZVP-rev2 and pob-TZVP-rev2. However, an optimization with the existing number of shells in these bases did not improve the quality of the resulting band structure. To remedy this, we added an additional 4d valence shell to the oxygen atom in the pob-TZVP-rev2-basis. In Fig.~\ref{fig:NiO} panel d), we illustrate that after optimizing the basis set in this updated format, the inconsistency for the unoccupied bands of the Gaussian basis set is fully removed when compared to the plane wave basis.

\subsection{Assessment of the basis set quality}\label{sec:basis_set_quality}
The difference in the eigenvalues between the plane wave and Gaussian basis set ${\Vert G - P - \Lambda \Vert}_{F}$ can be used as a criterion for a quick assessment of the basis quality in addition to the visual inspection of the differences. 
Here, we examine differences between the pob-DZVP-rev2 and pob-TZVP-rev2 bases for the NiO and MoS$_2$ solids studied previously, in addition to hexagonal BN and MnO solids. We show the band diagrams below for BN and MnO. the band diagrams for NiO and MoS$_2$ were presented in the previous section.

\subsubsection{BN}

\begin{figure*}%
\includegraphics[width=\textwidth]{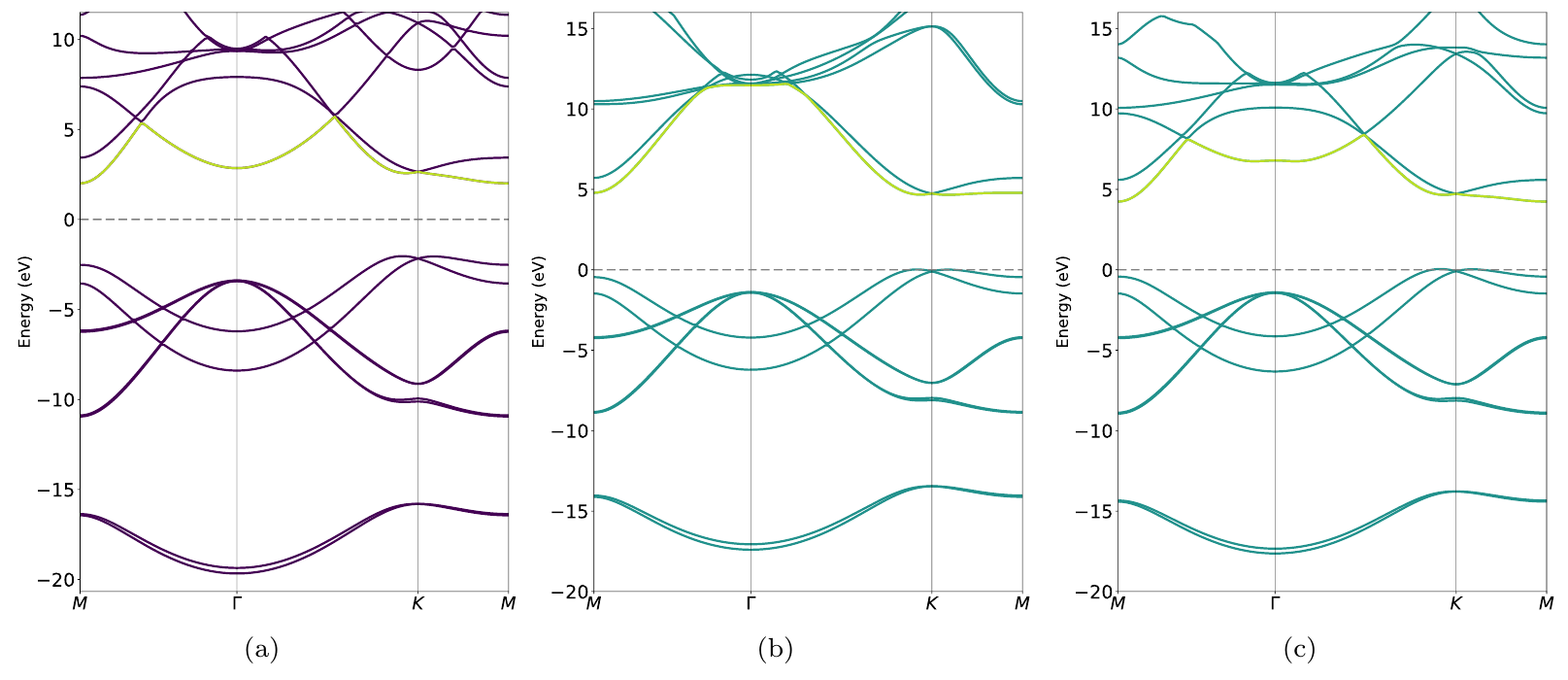}%
\caption{BN band structure plots:
(a) plane-wave basis calculation with cut off energy: 2000eV;
(b) Gaussian basis calculation with pob-DZVP-rev2-basis;
(c) Gaussian basis calculation with pob-TZVP-rev2-basis.}%
\label{fig:BN}%
\end{figure*}

For BN (Pearson symbol: hP4, crystal system: hexagonal, space group number: 194), as shown in Fig.~\ref{fig:BN}, the pob-DZVP-rev2-basis shows a qualitative discrepancy in the unoccupied part of the band structure near the Fermi level (that is marked in light green on the band structure diagram) when compared to the plane-wave reference calculations. The pob-TZVP-rev2-basis is qualitatively consistent with the reference. 

\subsubsection{MnO}
\begin{figure*}%
\includegraphics[width=\textwidth]{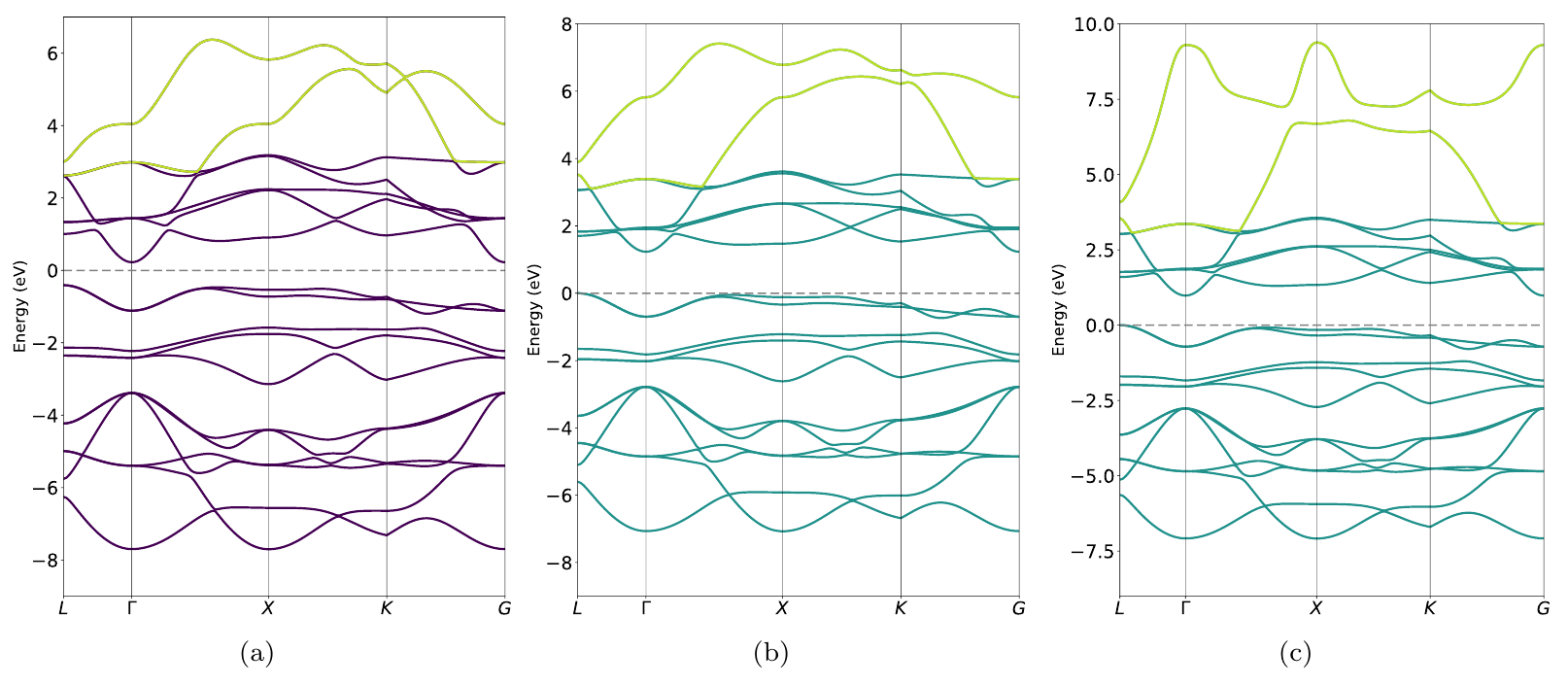}%
\caption{MnO band structure plots:
(a) plane-wave basis calculation with cut off energy: 2000eV;
(b) Gaussian basis calculation with pob-DZVP-rev2-basis;
(c) Gaussian basis calculation with pob-TZVP-rev2-basis.}%
\label{fig:MnO}%
\end{figure*}

For MnO (Pearson symbol: cF8, crystal system: fcc rocksalt, space group number: 225, AFM) the pob-DZVP-rev2 basis produces a better band structure result than the pob-TZVP-rev2-basis when compared with the plane wave reference. This indicates that the extra shells added in the TZVP basis are not essential or not optimized enough for a quantitatively correct description of the higher lying bands of MnO. This is illustrated in Fig.~\ref{fig:MnO}, where the incorrectly recovered bands are marked in light green. 

In Tab.~\ref{tab:mo_norm}, we show the MO eigenvalue norm  evaluated for pob-DZVP-rev2 and pob-TZVP-rev2 for the systems studied in this paper. We stress that this norm is evaluated for the valence orbitals and does not include core orbitals. The norms for the valence occupied and unoccupied orbitals are printed separately. 

From Tab.~\ref{tab:mo_norm}, we see that for BN there is a dramatic improvement once the basis set is increased. 

For NiO, the quality of the pob-DZVP-rev2 and pob-TZVP-rev2 is unchanged and is consistent with the band diagram plotted in Fig.~\ref{fig:NiO}.
However, our observation does not indicate that the total DFT energy evaluated in pob-TZVP-rev2 is higher than in the pob-DZVP-rev2 basis. The  pob-TZVP-rev2  energy is lower due to the optimization that happened for the core orbitals, while the valence bands remained mostly unaffected. 

For MnO, pob-TZVP-rev2 is worse for the unoccupied bands than pob-DZVP-rev2, according to the assessment using the difference of eigenvalues. This is consistent with the band diagram plotted in Fig.~\ref{fig:MnO}.

For MoS$_2$, the quality of pob-DZVP-rev2 and pob-TZVP-rev2 seems to be similar with some small worsening of the unoccupied part in the pob-TZVP-rev2 basis.

Subsequently, we also employ this criterion to analyze briefly and schematically the quality of the GTH basis~\cite{gth} for BN. GTH bases are another family of the systematically developed bases that are commonly used for solid state calculation. 

\begin{figure*}\label{fig:gth-BN}%
\includegraphics[width=\textwidth]{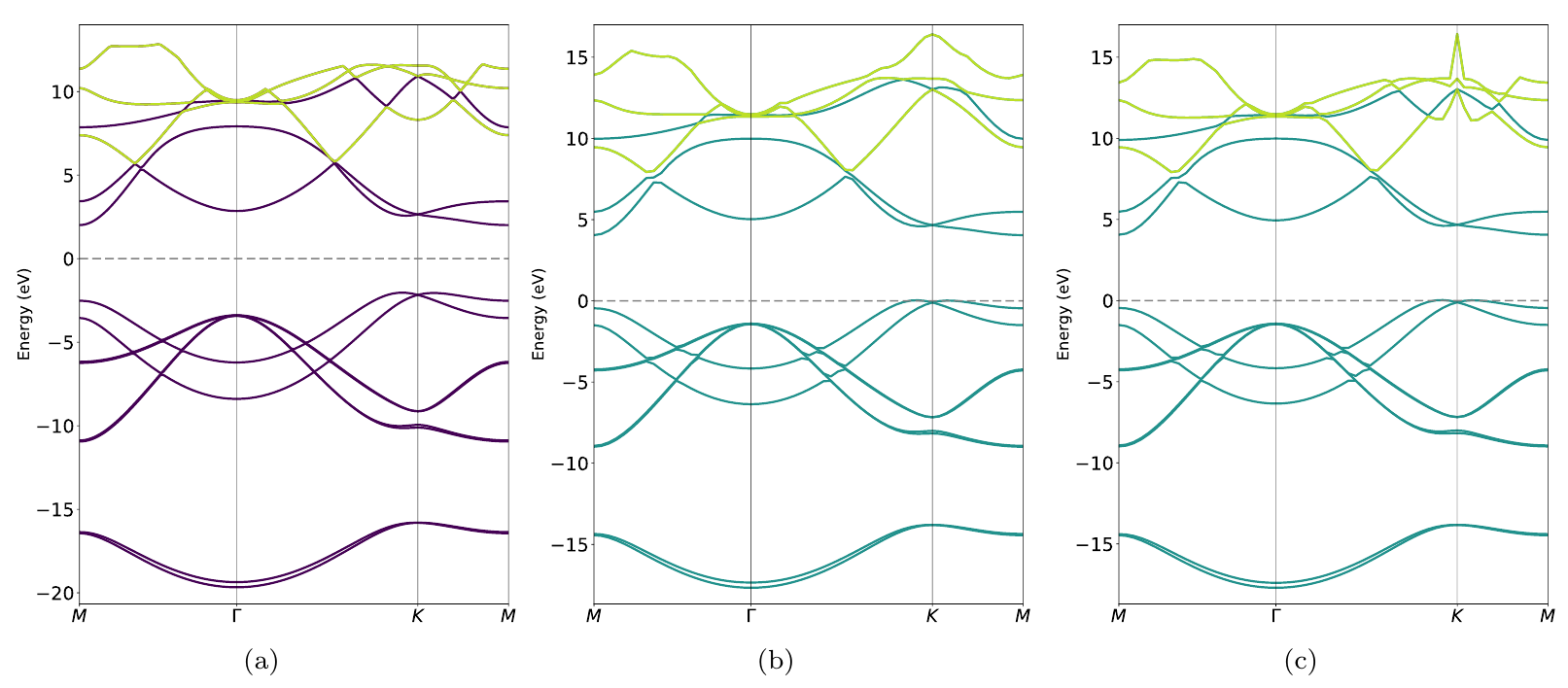}%
\caption{BN band structure plots calculated with GTH basis sets:
(a) plane-wave basis calculation with cut off energy: 2000eV;
(b) Gaussian basis calculation with DZVP-GTH basis;
(c) Gaussian basis calculation with TZVP-GTHP basis.}%
\end{figure*}

In the BN case, see Tab.~\ref{tab:bn_gth_mo_norm}, where both DZ and TZ bases are available, the gth-dzvp-molopt-sr basis with gth-pade as the pseudo-potential is yielding a similar deviation of the norm as the gth-tzvp-molopt-sr with the same pseudopotential. However, the maximal difference in the eigenvalues is larger for the unoccupied part of gth-tzvp-molopt-sr. This is consistent with the band diagram  illustrated in Fig.~\ref{fig:gth-BN}, where a spike in the unoccupied part of the band diagram arises for the gth-tzvp-molopt-sr basis.

\begin{center}
\begin{table}
\begin{tabular}{p{1.5cm} p{1.5cm} p{1.5cm} p{1.5cm} p{1.5cm}}
\hline
\hline
\hfil \multirow{2}{*}{System} & \multicolumn{2}{c}{OCC} & \multicolumn{2}{c}{UNOCC}\\
{}  & \hfil \footnotesize{DZVP}   & \hfil \footnotesize{TZVP}    & \hfil \footnotesize{DZVP}   & \hfil\footnotesize{TZVP} \\
\hline
\hline
\hfil BN
        &  \multicolumn{1}{c}{0.7186} 
        &  \multicolumn{1}{c}{0.2096} 
        &  \multicolumn{1}{c}{10.7039} 
        &  \multicolumn{1}{c}{4.2203}\\
    
\hline
\hfil NiO 
        &  \multicolumn{1}{c}{0.5911} 
        &  \multicolumn{1}{c}{0.7998} 
        &  \multicolumn{1}{c}{5.9098} 
        &  \multicolumn{1}{c}{5.9848}\\
\hline
\hfil MnO 
        &  \multicolumn{1}{c}{0.7328} 
        &  \multicolumn{1}{c}{0.7906} 
        &  \multicolumn{1}{c}{1.6521} 
        &  \multicolumn{1}{c}{2.3830}\\
\hline
\hfil MoS$_2$ 
        &  \multicolumn{1}{c}{0.8715} 
        &  \multicolumn{1}{c}{0.8703} 
        &  \multicolumn{1}{c}{0.6222} 
        &  \multicolumn{1}{c}{0.7930}\\
\hline
\hline
\end{tabular}
\caption{MO eigenvalue norms calculated with respect to plane-wave basis calculations. Basis sets used are pob-DZVP-rev2 and pob-TZVP-rev2, respectively. BN: 8 occupied bands and 6 unoccupied bands are used; NiO: 22 occupied bands and 6 unoccupied bands are used; MoS$_2$: 10 occupied bands and 6 unoccupied bands are used; MnO: 11 occupied bands and 6 unoccupied bands are used. } \label{tab:mo_norm}
\end{table}
\end{center}

\begin{table}
\begin{tabular}{p{1.5cm} p{1.5cm} p{1.5cm} p{1.5cm} p{1.5cm}}
\hline
\hline
\hfil \multirow{2}{*}{Basis} &  \multicolumn{2}{c}{ Norm} & \multicolumn{2}{c}{Max. Diff}\\
{}   & \hfil \footnotesize{OCC}   & \hfil \footnotesize{UNOCC}   & \hfil \footnotesize{OCC}   & \hfil \footnotesize{UNOCC} \\
\hline
\hline
\hfil DZVP &  \hfil 0.2260 & \hfil 3.0873   & \hfil 2.0691  & \hfil 4.8523\\
\hline
\hfil TZVP &  \hfil 0.2432 & \hfil 3.0329  & \hfil 2.0668   & \hfil 4.8795\\
\hline
\hline
\end{tabular}
\caption{BN MO eigenvalue norm calculated with respect to plane-wave basis calculations. Basis sets used are DZVP-GTH and TZVP-GTH, respectively, with pseudo-potential GTH-PADE. 8 occupied bands and 6 unoccupied bands are used in the norm calculation.}\label{tab:bn_gth_mo_norm}
\end{table}

\section{Conclusions}\label{sec:conclusions}
We have presented a Gaussian basis set optimization scheme for solid state calculations. This optimization scheme results in the minimization of the total DFT energy, the condition number of the overlap matrix,  as well as minimization of the difference between the eigenvalues evaluated in the reference plane wave basis and the currently optimized Gaussian basis set.

Our aim of designing such a material specific optimization scheme is to employ the moderately sized optimized bases in correlated periodic {\em ab-initio} calculations such as second-order Green's function theory (GF2)~\cite{Dahlen05,Zgid14,Phillips15,Rusakov14,Rusakov16,Welden16,Kananenka_grid_16,Kananenka16,kananenka_hybrif_gf2,Iskakov_Chebychev_2018}, GW~\cite{Hedin65,Aryasetiawan98,QPGW_Schilfgaarde,Stan06,Koval14,PhysRevB.66.195215,GW100,Holm98,Tran_GW_SEET}, and self-energy embedding theory (SEET)~\cite{Kananenka15,Zgid17,Rusakov19,Iskakov20} in order to evaluate and interpret spectral functions of solids. Gaussian bases are very useful for the physical interpretation of the peaks present in spectral functions and for assigning them to electron attachment or removal processes from $s$, $p$, or $d$ orbitals.  Since these {\em ab-initio} methods are computationally demanding, a moderately sized Gaussian basis is extremely advantageous in carrying out these calculations and comparing them to experiment.

We believe that this scheme will be particularly useful for periodic calculations, where a given element may display vastly different properties depending on the surrounding crystal lattice environment.
A significantly changed basis may then capture the changes in the valence orbitals arising from different physical/chemical properties of the crystal, while remaining of moderate size. 

The improvement of basis sets quality for crystalline systems is not as straightforward as for molecular ones, for multiple reasons. 
We have shown in this work that for some compounds simply increasing basis set size by following the typical double, triple, and quadruple zeta hierarchy (eg. using a pob-TZVP-rev2 instead of pob-DZVP-rev2 for NiO) may not necessarily lead to an improvement of the basis set quality without additional material specific optimization.
Moreover, the onset of linear dependence issues in periodic systems may be delayed by having a material specific optimization, thereby allowing one to employ bases of moderate sizes that are geared towards recovering experimental results.
The advantage of the scheme presented here is the possibility of building a basis set of user desired quality by insisting that band after band is optimized when compared to a plane wave calculation. Alternatively, if so desired by a user, a series of bands within a given energy window can be optimized.

The obvious drawback of the scheme proposed here is a possible lack of transferability between different compounds. However, our main motivation of designing this scheme is to use it for {\em ab-initio} post-DFT calculations of strongly correlated compounds. For these materials, few basis sets are readily available thus an optimization is frequently necessary before starting calculations. Moreover, for some of the materials of interest the experimental data may be scarce and calculations should have predictive quality. In such cases, material specific optimization is helpful to ensure that even a moderately sized basis set is the most optimal for the calculated material and all possible differences when comparing against experimental data come for the computational method employed.
Finally, we believe that as post-DFT, correlated, {\em ab-initio} calculations for solids in Gaussian bases will become more prevalent, a database of material specific bases can be created, thus alleviating the issue of transferability. 

At last, we reiterate that the scheme presented here is not only useful for basis set optimization but also for a quick assessment of the basis set quality as illustrated in Sec.~\ref{sec:basis_set_quality}. In such a case, a score based on the eigenvalue differences between the plane wave and Gaussian basis can be assigned in order to aid the visual inspection necessary for the assessment of the basis set quality.

\acknowledgments
Y. Z. and D. Z. acknowledge support from NSF grant CHE-1453894.
E.G. was sponsored by the Simons foundation via the
Simons Collaboration on the Many-Electron Problem.
\bibliographystyle{apsrev4-1}
\bibliography{bib.bib}
%

\end{document}